\documentclass{article}
\usepackage{tabularx,booktabs,multirow,delarray,array}
\usepackage{graphicx,amssymb,amsmath,amssymb}
\usepackage{latexsym}
\usepackage{fullpage}
\usepackage{cite}
\usepackage{epsfig}
\usepackage{subcaption}
\usepackage[utf8]{inputenc}
\usepackage[export]{adjustbox}
\usepackage{wrapfig}
\usepackage{ragged2e}
\usepackage{pifont}
\usepackage{multirow}
\usepackage{amssymb}
\usepackage{makecell}
\usepackage{multicol}
\usepackage{hyperref}
\hypersetup{
  colorlinks   = true, %Colours links instead of ugly boxes
  urlcolor     = black, %Colour for external hyperlinks
  linkcolor    = red, %Colour of internal links
  citecolor   = red %Colour of citations
}

\title{Character Animation in AR: Character Animation in AR: a mobile application development study}

\author{Sukanya Bhattacharjee\\ %
\scriptsize
Department of Computer Science and Engineering\\
\scriptsize IIT Bombay\\
\scriptsize sukanyabhat@cse.iitb.ac.in
\and Parag Chaudhuri\\ %
\scriptsize
Department of Computer Science and Engineering\\
\scriptsize IIT Bombay\\
\scriptsize paragc@cse.iitb.ac.in
} %
\date{}

\begin{document}

\maketitle

% \section{Interactive authoring in AR}\label{interactAR}
Digital preservation of the cultural heritages is one of the major applications of various computer graphics and vision algorithms. The advancement in the AR/VR technologies is giving the cultural heritage preservation an interesting spin due to its immense visualization ability. The use of these technologies to digitally recreate heritage sites and art is becoming a popular trend. A project, called Indian Digital Heritage (IDH), for recreating the heritage site of Hampi, Karnataka during Vijaynagara empire ($1336$ - $1646$ CE) has been initiated by the Department of Science and Technology (DST) few years back. Immense work on surveying the site, collecting geographical and historic information about the life in Hampi, creating 3D models for buildings and people of Hampi and many other related tasks has been undertaken by various participants of this project. A major part of this project is to make tourists visiting Hampi visualize the life of people in ancient Hampi through any handy device. With such a requirement, the mobile AR based platform becomes a natural choice for developing any application for this purpose. We contributed to the project by developing an AR based mobile application to recreate a scene from Virupaksha Bazaar of Hampi with two components - author scene with augmented virtual contents at any scale, and visualize the same scene by reaching the physical location of augmentation. We develop an interactive application for the purpose of digitally recreating ancient Hampi. Though the focus of this work is not creating any static or dynamic content from scratch, it shows an interesting application of the content created in a real world scenario.

We dedicate Section~\ref{mobFrame} to describe various tests that we did to decide upon the suitable AR framework to be used. Section~\ref{inAuth} discusses the features and implementation details for creating the interactive authoring application. The conceptual description of the idea behind the user application is discussed in Section~\ref{userView}. The final results of both the applications are presented in Section~\ref{hamRes}.

\section{Mobile AR framework tests \label{mobFrame}}
\begin{figure}[h!t]
    \begin{subfigure}{0.5\textwidth}
        \includegraphics[width=\linewidth]{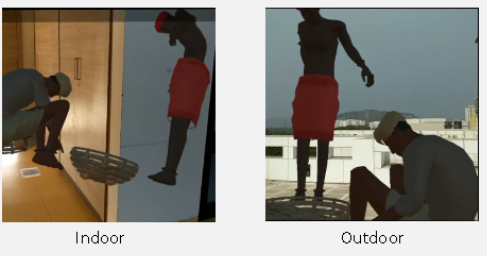} 
        \caption{Vuforia}
        \label{fig41a}
    \end{subfigure}
        \begin{subfigure}{0.5\textwidth}
        \includegraphics[width=\linewidth]{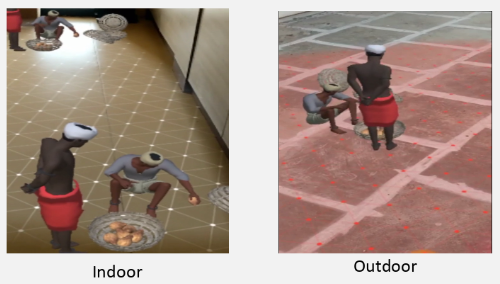}
        \caption{Google ARCore}
        \label{fig42a}
    \end{subfigure}
    \caption{Stability and distance tests for AR frameworks}
    \label{fig4}
\end{figure}

The two major requirements with respect to the AR platform needed to develop the desired application is stability of augmentation under camera movement and augmentation from a fairly large distance. Vuforia \cite{vuforia} and Google ARCore \cite{arcore} are two popular commercially available AR platforms. We evaluated these two platforms with respect to these two requirements to make choice of the framework to be used. The tests are conducted under the following conditions:

\begin{itemize}
    \item[-] Both the frameworks are tested with the same virtual objects to be augmented.
    \item[-] Both the frameworks are tested under exactly the same indoor and outdoor conditions (i.e. lighting condition, geographical area, etc.)
\end{itemize}

We attempt to augment a set animated characters to a planar area both indoors and outdoors by tapping the screen of the mobile phone. Figure \ref{fig4} shows the results of the corresponding tests for each of the framework.

In case of Vuforia framework, it is marker based and limited by the visibility of the marker, and the augmented content is prone to more flickering. In case of ARCore framework, it is markerless and the detected plane automatically extends as the camera moves, and the tracking is quickly re-acquired. Since a more stable augmentation is provided by the ARCore framework, we choose it as a platform for developing the required application along with Unity 3D \cite{unity} as the development environment. All the 3D animated models of people is created using Blender \cite{blender}.

\section{Interactive Authoring \label{inAuth}}
In this section, we describe the first component of the application which is an interactive authoring application to place different sets of animated characters at various physical locations.
The goal of this application is creating an augmented scene with various life size characters with the ability to store their locations. We call the characters as \textit{prefabs}, which is a Unity specific terms. For this, we implemented the following five features which can be used to achieve the above goal.

\begin{figure}[h!t]
    \begin{subfigure}{0.5\textwidth}
        \includegraphics[width=\linewidth]{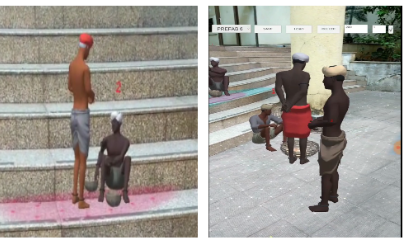} 
        \caption{Placing prefabs}
        \label{fig51a}
    \end{subfigure}
        \begin{subfigure}{0.5\textwidth}
        \includegraphics[width=\linewidth]{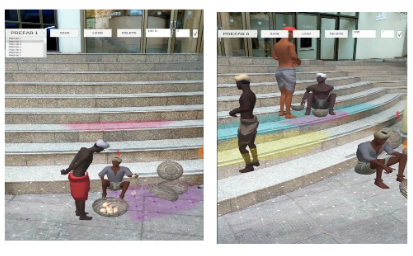}
        \caption{Selecting prefabs dynamically}
        \label{fig52a}
    \end{subfigure}
    \caption{Placing dynamically selected prefabs}
    \label{fig5}
\end{figure}

\textbf{Placing prefabs} is the core feature of this application which is achieved by tapping on any point on the mobile screen which corresponds a point contained in the auto-detected physical plane (see Figure \ref{fig51a} for an example). The plane detection is a part of the internal functioning of ARCore in form of \texttt{Trackables} which helps to keep track of the environment. The prefabs are anchored to the planes where they are placed to keep their positions fixed throughout all the frames.

\begin{figure}[h!t]
    \begin{subfigure}{0.5\textwidth}
        \includegraphics[width=\linewidth]{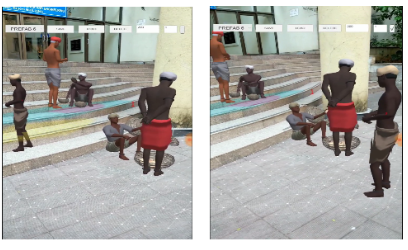} 
        \caption{Relocating prefabs}
        \label{fig61a}
    \end{subfigure}
        \begin{subfigure}{0.5\textwidth}
        \includegraphics[width=\linewidth]{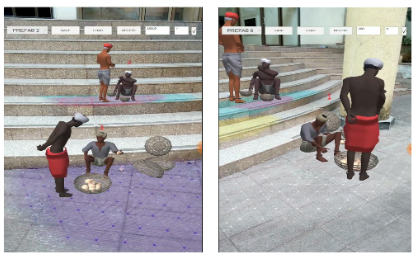}
        \caption{Scaling prefabs}
        \label{fig62a}
    \end{subfigure}
    \caption{Editing augmented prefabs}
    \label{fig6}
\end{figure}

\begin{figure}[h!t]
    \begin{subfigure}{0.5\textwidth}
        \includegraphics[width=\linewidth]{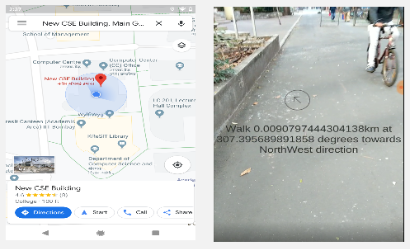} 
        \caption{Prototype navigation system}
        \label{fig71a}
    \end{subfigure}
        \begin{subfigure}{0.5\textwidth}
        \includegraphics[width=\linewidth]{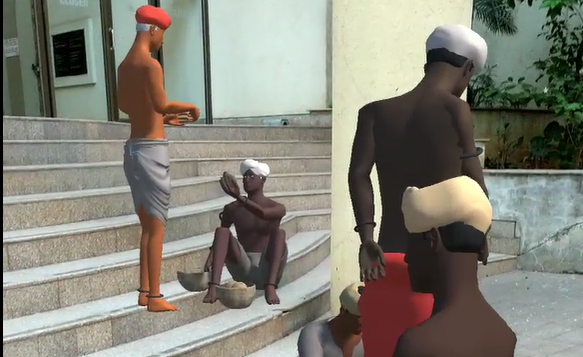}
        \caption{User view}
        \label{fig72a}
    \end{subfigure}
    \caption{User view of the scene}
    \label{fig7}
\end{figure}

\textbf{Dynamic selection} of prefabs is a feature which allows placing of a large number of different sets of characters (see Figure \ref{fig52a} for example). This selection is provided by a dropdown menu showing all prefabs as different options.

\textbf{Relocating} prefabs is important for providing ease of usage to the user authoring the scene. It helps the user to change location of a placed prefab as per his or her need at some later stage of authoring. This feature is provided by using identification number of a prefab(which is associated with a prefab when it is augmented). The relocation works in a two step manner - the prefab corresponding to the selected identification number is deleted and then the same prefab can be placed at a different location using the \texttt{placing prefab} feature (see Figure \ref{fig61a} for example). This feature includes the deletion feature inherently.

\textbf{Scaling} prefabs is a feature which helps the user to create life-size character augmentations in the scene. This makes the augmentations scalable from a miniature scale to a life-size scale. This feature is provided by using the identification number of a prefab and providing a scale value for it (see Figure \ref{fig62a} for example).

\textbf{Saving the augmentations} is another core feature for this application which is useful for the second component of our application, the user view application. For this purpose, the user authoring the scene just needs to use the \textit{save} GUI button. Our interface stores the physical location, scale and identification number of each augmented character which is made available to the user application. 

\section{User view \label{userView}}
This section is dedicated for discussing the second component of the application, the user view application. The purpose of this application is to make the user of this view the augmented characters at the same location with their devices. But the ARCore framework works on a local coordinate system and, thus, the locations that we store during the authoring process are relative to the origin during that session. So, using the stored locations as it is does not work. To curb this issue, we propose to put a GPS based navigation layer on top the viewing application. The navigation layer helps the user to reach the vicinity of the augmentations. To compensate for change in the camera rotation as compared to that in case of the authoring application, our interface navigates the user to line of sight of the augmented locations. Once the user is at the appropriate location, the set of characters saved in the authoring stage are anchored to their save location. Figure \ref{fig71a} shows a prototype of a possible navigation system which uses a mix of Google Maps \cite{gmap} and Unity's GPS system for better precision. Figure \ref{fig72a} shows an example view which the user can experience in the augmentation location.

\begin{figure}[h!t]
\begin{tabular}{cc}
  \includegraphics[width=0.5\textwidth]{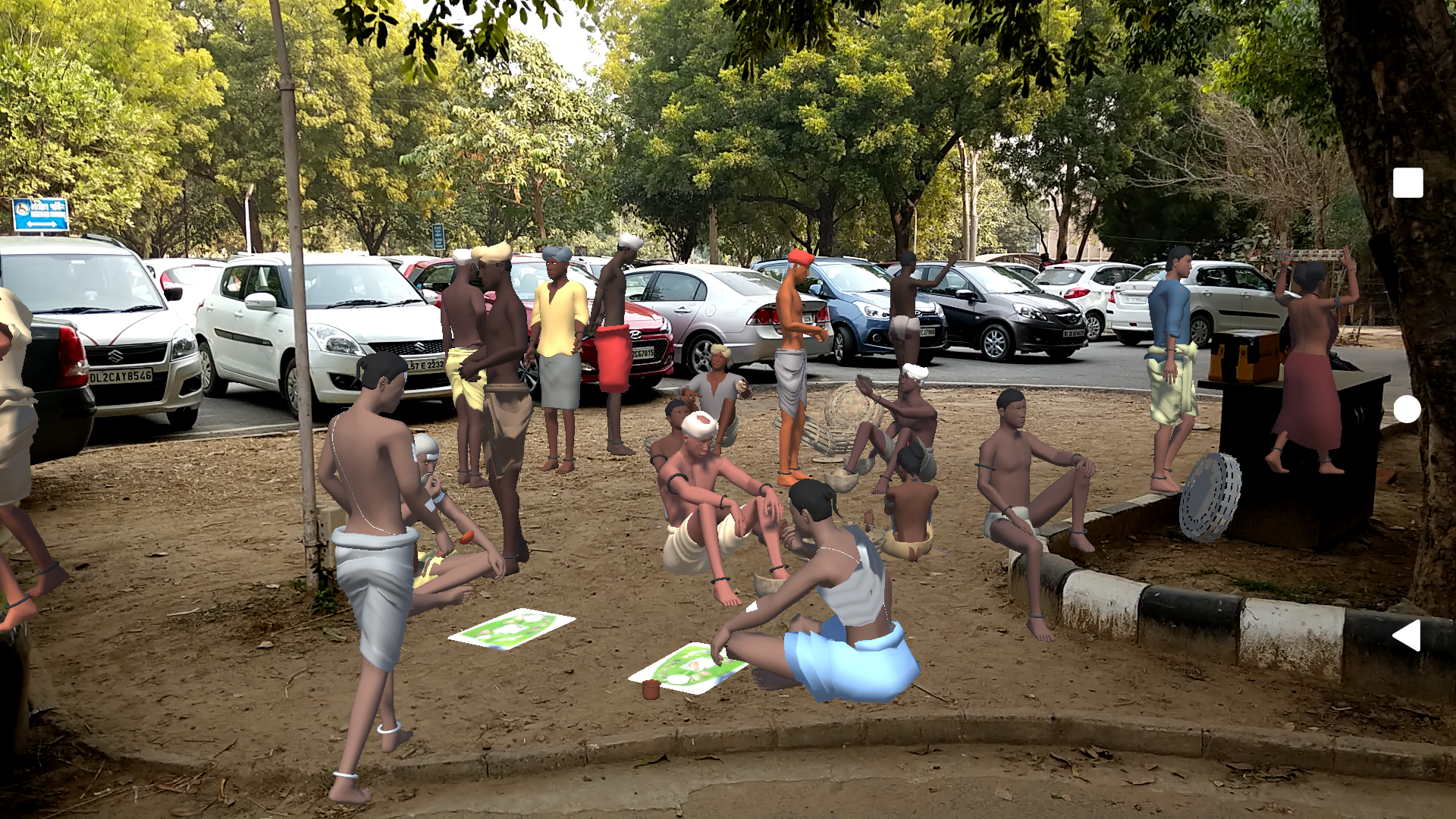} & 
  \includegraphics[width=0.5\textwidth]{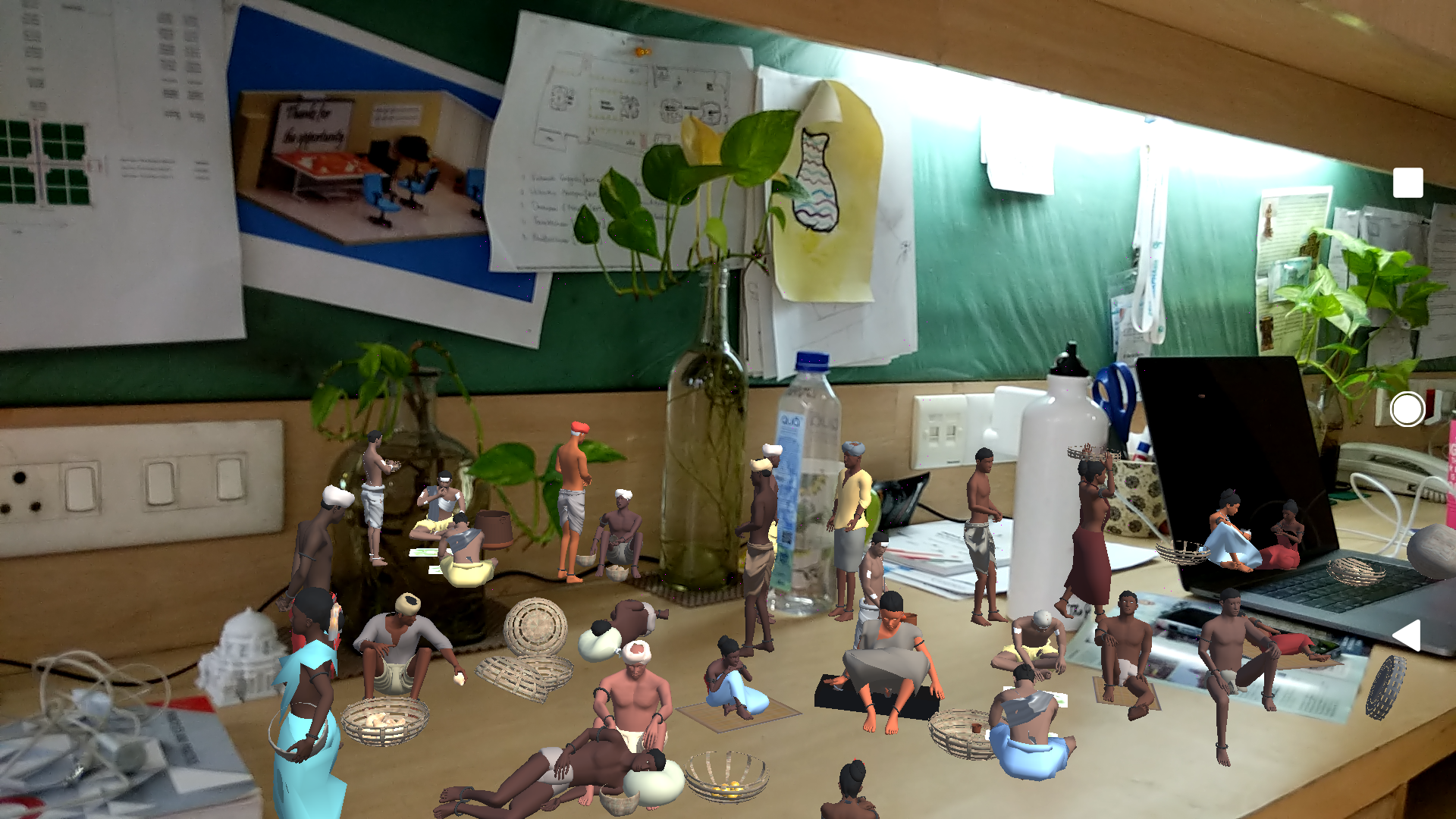} \\
  \includegraphics[width=0.5\textwidth]{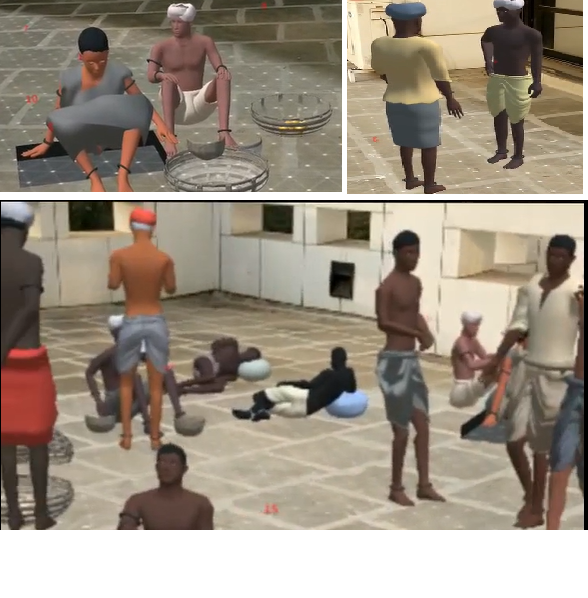} &
  \includegraphics[width=0.5\textwidth]{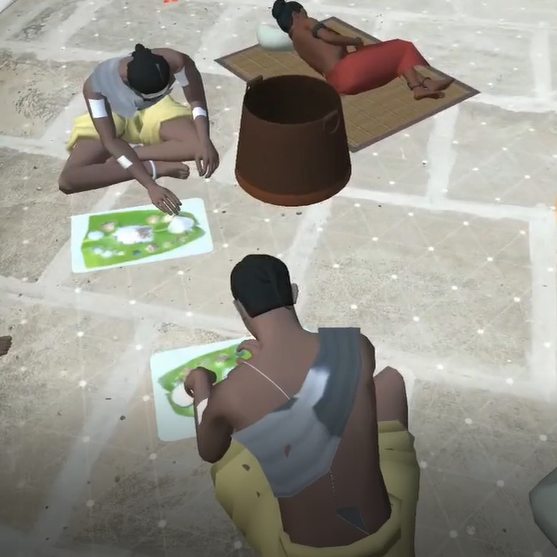} \\
\end{tabular}
\caption{Scenes authored by our interface in different condition}
\label{fig8}
\end{figure}

\section{Results \label{hamRes}}
We present few snapshots of the results that our interfaces produces at different conditions in the Figure \ref{fig8}. All our tests are done with the Google Pixel 2 phone. This application can be used for authoring any augmented scene with any set of characters. Therefore, this application can be used for other purposes like creating live demonstration for teaching some concepts like solar systems or historic event in a classroom set up as well. It can also find its usage in demonstration of some medical procedures as well. The only requirement for these would be to have 3D assets or prefabs which form the scene.

\vspace{10mm}

\bibliographystyle{plain}
\bibliography{sem_ref}

\begin{thebibliography}{1}

\bibitem{blender}
Blender.
\newblock Blender 2.79b, 2017.
\newblock \url{https://www.blender.org/} Last accessed on 15-10-2019.

\bibitem{arcore}
Google.
\newblock A{RC}ore, 2020.
\newblock \url{https://developers.google.com/ar} Last accessed on 08-02-2020.

\bibitem{gmap}
Google.
\newblock Google {M}aps, 2020.
\newblock \url{https://developers.google.com/maps/documentation} Last accessed on 09-02-2020.

\bibitem{vuforia}
P{TC}.
\newblock Vuforia, 2020.
\newblock \url{https://developer.vuforia.com/} Last accessed on 08-02-2020.

\bibitem{unity}
Unity Technologies.
\newblock Unity 3{D}, 2020.
\newblock \url{https://unity.com/} Last accessed on 08-02-2020.

\end{thebibliography}
\end{document}